\documentclass[12pt,notoc]{JHEP3} 

\usepackage{graphicx}
\usepackage{dcolumn}
\usepackage{bm}

\newcommand{\eps}{\epsilon}

\def\cA{{\cal A}}
\newcommand{\beq}{\begin{equation}}
\newcommand{\eeq}{\end{equation}}
\newcommand{\beqa}{\begin{eqnarray}}
\newcommand{\eeqa}{\end{eqnarray}}

\def\Gr{{\rm Gr}}

\def\OP{{\rm OP}}
\def\cg{c_\Gamma}

\def\N{\cal{N}}
\def\cA{\cal A}
\def\Atree{A^{\rm tree}}
\def\tree{\mbox{\rm tree}}
\def\Tr{\mathop{\rm Tr}\nolimits}

\def\za{\cal{N}}
\def\za#1.#2{\left\langle#1 \hskip .15 mm #2\right\rangle}

\title{The one-loop amplitude for six-gluon scattering}

\author{R.~K.~Ellis\footnote{email: ellis@fnal.gov}\\
  Fermilab, Batavia, IL 60510, USA \\
  Physics Department, Theory Division, CERN, 1211 Geneva 23,
  Switzerland}

\author{W.~T.~Giele\footnote{email: giele@fnal.gov}\\
  Fermilab, Batavia, IL 60510, USA}

\author{G.~Zanderighi\footnote{email: giulia.zanderighi@cern.ch}\\
  Physics Department, Theory Division, CERN, 1211 Geneva 23,
  Switzerland}

\abstract{We present results for the six-gluon scattering amplitude at
  one-loop.  Since our method is semi-numerical, it yields the result
  for arbitrary momenta and helicities of the external gluons.  We
  evaluate the colour-ordered sub-amplitudes with gluons, fermions and
  scalars running in the internal loop.  This is more than sufficient
  to give a complete description of six-gluon scattering at one-loop
  in QCD. Combination of these results into amplitudes with
  ${\cal{N}}=4$ and ${\cal{N}}=1$ multiplets of supersymmetric
  Yang-Mills or with a complex scalar in the internal loops allows
  comparison with analytic results in the literature.  The numerical
  results for most of the helicity combinations with loops of complex
  scalars are new.}

\date{\today}

\preprint{CERN-PH-TH/2006-032, FERMILAB-PUB-06-031-T}

\begin{document}

\section{Introduction}
Phenomenology at the LHC often involves high multiplicity final
states.  For example, backgrounds to Higgs searches involve processes
such as $PP\rightarrow W^+W^- + 2~\mbox{jets}$ and $PP\rightarrow
t\bar{t}\ +\ b\bar{b}$. Both these examples involve $2\rightarrow 4$
scatterings. At leading order (LO) such high multiplicity final state
amplitudes can be evaluated using either numerical recursive
techniques~\cite{Berends:1987me,Mangano:2002ea,Draggiotis:2002hm} or
other numerical and/or algebraic
techniques~\cite{Ishikawa:1993qr,Stelzer:1994ta,Krauss:2001iv,Maltoni:2002qb,Boos:2004kh}.

However, ${\cal O}\left(\alpha_S\right)$, next-to-leading order (NLO)
corrections to the scattering amplitudes are desirable. Not only do
NLO corrections give a first reliable prediction of total rates, they
also give a good error estimate on the shapes of distributions.  At
NLO the current state of the art for hadron colliders are $2
\rightarrow 3$ processes.
Thus NLO predictions for $PP\rightarrow 3\ 
\mbox{jets}$~\cite{Kilgore:1996sq,Nagy:2003tz} (based on virtual
corrections of ref.~\cite{Bern:1993mq,Bern:1994fz,Kunszt:1994tq}) and
$PP\rightarrow V\ +\ 2\ \mbox{jets}$~\cite{Campbell:2002tg} (based on
virtual corrections of
ref.~\cite{Glover:1996eh,Campbell:1997tv,Bern:1997sc}) are known, and
codes for $PP\rightarrow t\bar{t}\ +\ 
\mbox{jet}$~\cite{Brandenburg:2004fw,Uwer:2005tq} and $PP\rightarrow
H\ +\ 2\ \mbox{jets}$ via gluon fusion~\cite{Ellis:2005qe} are under
construction. Other processes such as $PP\rightarrow V_1V_2\ +\ 
\mbox{jet}$ and $PP\rightarrow V_1V_2V_3$ are now feasible.

By contrast the consideration of $2\rightarrow 4$ processes is still
in its infancy.  In electroweak physics the full one-loop electroweak
corrections to $e^+e^-\rightarrow\ 4\ \mbox{fermions}$ were calculated
in Ref.~\cite{Denner:2005fg,Denner:2005nd}. However the calculation of
NLO $2\rightarrow 4$ QCD scattering cross sections is currently
unexplored.  Such a calculation involves both the evaluation of the
one-loop six-point virtual corrections and the inclusion of the
$2\rightarrow 5$ scattering bremsstrahlung contributions through Monte
Carlo integration.

In this paper we consider the virtual corrections to six-gluon
scattering which is relevant for a calculation of $PP\rightarrow
4$~jets.  By considering the one-loop corrections to $gg\rightarrow
gggg$ we select the most complicated QCD six-point processes. If the
amplitude is calculated in terms of Feynman diagrams, the number of
diagrams is very large and the gauge cancellations between these
diagrams is the most severe.  These cancellations could be a concern
in a semi-numerical procedure; the six-gluon amplitude therefore
provides a stringent test of the method.  In this paper we consider
neither the bremsstrahlung contributions, nor the one-loop processes
involving external quarks, which are needed to obtain results for a
physical cross section.

The technique for the analytic calculation of the one-loop corrections
to multi-gluon amplitudes which is relevant for this paper 
is the decomposition of the calculation into simpler
pieces with internal loops of ${\cal{N}}=4$ and ${\cal{N}}=1$
multiplets of super-symmetric Yang-Mills particles and a residue
involving only scalar particles in the
loops~\cite{Bern:1993mq,Bern:1994zx,Bern:1994cg}.
After recent
advances~\cite{Bidder:2004tx,Bern:2005ji,Bern:2005cq,Britto:2005ha},
all supersymmetric contributions have been computed analytically,
however not all of the scalar contributions for six-gluon amplitudes
(or higher) are known yet. We present here numerical results for
six-gluon contributions.  For supersymmetric pieces we provide
completely independent cross-checks of analytical results.

Although all one-loop $2\rightarrow 2$ and almost all of the currently
known $2\rightarrow 3$ amplitudes were calculated using analytic
techniques, we believe that semi-numerical or hybrid
numerical/analytic techniques offer promise for more rapid progress.
This technique was demonstrated recently for the case of the one-loop
$\mbox{H}\ +\ 4\ \mbox{partons}$ amplitude~\cite{Ellis:2005qe}.

Many methods have been proposed to calculate NLO amplitudes, both
semi-numerical~\cite{Fleischer:1999hq,Passarino:2001jd,Binoth:2003ak,Duplancic:2003tv,
  Nagy:2003qn,Belanger:2003sd,delAguila:2004nf,Denner:2002ii,Giele:2004iy,vanHameren:2005ed,Denner:2005nn}
or numerical~\cite{Soper:2001hu,Anastasiou:2005cb}.  Of these methods
only a few have actually been used to evaluate one-loop amplitudes.
Only by using the methods in explicit calculations one can be sure
that all numerical issues have been addressed properly.

In section II we discuss the colour algebra involved with the
evaluation of a six-gluon amplitude. The numerical techniques used in
this paper are discussed in section III, while in section IV the
comparison is made with numerous super-symmetric and the few scalar
results, which exist in the literature. Finally, our conclusions in
section V summarize the paper.

\section{Six-gluon amplitude at one-loop}
At tree-level, amplitudes with $n$ external gluons can be decomposed
into colour-ordered sub-amplitudes, multiplied by a trace of $n$
colour matrices, $T^a$.  The traceless, hermitian, $N_c\times N_c$
matrices, $T^a$, are the generators of the $SU(N_c)$ algebra.
Following the usual conventions for this branch of the QCD literature,
they are normalized so that $\Tr( T^a T^b) = \delta^{ab}$.  Summing
over all non-cyclic permutations the full amplitude ${\cA}^{\rm
  \scriptsize \tree}_ n$ is reconstructed from the sub-amplitudes
$A_n^{\rm \scriptsize
  \tree}(\sigma)$~\cite{Berends:1987me,Mangano:1987xk},
\begin{equation}
{\cA}_{n}^{\rm tree}(\{p_i,\lambda_i,a_i\}) =
g^{n-2} \sum_{\sigma \in S_n/Z_n} \Tr( T^{a_{\sigma(1)}}
\cdots T^{a_{\sigma(n)}} )
\ \Atree_n (p_{\sigma(1)}^{\lambda_{\sigma(1)}},\ldots,
            p_{\sigma(n)}^{\lambda_{\sigma(n)}})\ .
\end{equation}
The momentum, helicity ($\pm$), and colour index of the $i$-th
external gluon are denoted by $p_i$, $\lambda_i$, and $a_i$
respectively.  $g$ is the coupling constant, and $S_n/Z_n$ is the set
of $(n-1)!$ non-cyclic permutations of $\{1,\ldots, n\}$.

The expansion in colour sub-amplitudes is slightly more complicated at
one-loop level.  Let us consider the case of massless internal
particles of spin $J=0,1/2,1$ corresponding to a complex scalar, a
Weyl fermion or a gluon.  If all internal particles belong to the
adjoint representation of SU$(N_c)$, the colour decomposition for
one-loop $n$-gluon amplitudes is given by~\cite{Bern:1990ux},
\begin{equation}
{\cal A}_n^{[J]} ( \{p_i,h_i,a_i\} ) = g^n
      \sum_{c=1}^{\lfloor{n/2}\rfloor+1}
      \sum_{\sigma \in S_n/S_{n;c}}
     \Gr_{n;c}( \sigma ) \,A_{n;c}^{[J]}(\sigma) \,,
\label{Oneloopform}
\end{equation}
where ${\lfloor{x}\rfloor}$ denotes the largest integer less than or
equal to $x$ and $S_{n;c}$ is the subset of $S_n$ which leaves the
double trace structure in $\Gr_{n;c}(1)$ invariant.
The leading-colour structure is simply given by,
\begin{equation}
\Gr_{n;1}(1) = N_c\ \Tr (T^{a_1}\cdots T^{a_n} ) \,.
\end{equation}
The subleading-colour structures are given by products of colour traces
\begin{equation}
\Gr_{n;c}(1) = \Tr( T^{a_1}\cdots T^{a_{c-1}} )\,
\Tr ( T^{a_c}\cdots T^{a_n}) \,.
\end{equation}

The subleading sub-amplitudes $A_{n;c>1}$ are determined by the
leading ones $A^{[1]}_{n;1}$ through the merging
relation~\cite{Kleiss:1988ne,Bern:1990ux,Bern:1994zx,DelDuca:1999rs}
\begin{equation}
A^{[1]}_{n;c>1}(1,2,\ldots,c-1;c,c+1,\ldots,n)\ =\
 (-1)^{c-1} \sum_{\sigma \in \OP\{\alpha\}\{\beta\}}
 A^{[1]}_{n;1}(\sigma_1,\ldots,\sigma_n) \, ,
\label{Kleiss-Kuijf}
\end{equation}
where $\alpha_i \in \{\alpha\} \equiv \{c-1,c-2,\ldots,2,1\}$,
$\beta_i \in \{\beta\} \equiv \{c,c+1,\ldots,n-1,n\}$, and
$\OP\{\alpha\}\{\beta\}$ is the set of ordered permutations of
$\{1,2,\ldots,n\}$ but with the last element $n$ fixed.  The ordered
permutations are defined as a set of all mergings of $\alpha_i$ with
respect to the $\beta_i$, such that the cyclic ordering of the
$\alpha_i$ within the set $\{\alpha\}$ and of the $\beta_i$ within the
set $\{\beta\}$ is unchanged. In practice, since $n$ is fixed, no
further cycling of the set $\{\beta\}$ is required.  Thus a complete
description can be given in terms of the leading colour sub-amplitudes
$A_{n;1}$ alone.

The contribution of a single flavour of Dirac fermion
in the fundamental representation, (relevant for quarks in QCD) is
\begin{equation}
{\cA}_{n}^{\rm Dirac}(\{p_i,\lambda_i,a_i\}) =
g^n \sum_{\sigma \in S_n/Z_n} \Tr( T^{a_{\sigma(1)}}
\cdots T^{a_{\sigma(n)}} )
\ A^{[1/2]}_{n;1} (p_{\sigma(1)}^{\lambda_{\sigma(1)}},\ldots,
            p_{\sigma(n)}^{\lambda_{\sigma(n)}})\ .
\end{equation}
Simple colour arguments~\cite{Bern:1990ux} allow one to demonstrate
that this colour sub-amplitude is the same as the leading colour
sub-amplitude for a single Weyl fermion in the adjoint representation
defined in Eq.~(2.2).

Since the subleading colour amplitudes are not independent, we shall
henceforth drop them from our discussion. To simplify the notation we
shall also drop the subscripts $n$ and $c$. The amplitude denoted by
$A$ will thus refer to leading colour amplitude with six external
gluons.

\section{Method of calculation}
The method we use is purposely kept as simple as possible.  Especially
in numerical methods this is desirable for both keeping track of
numerical accuracy and code transparency.

To generate all the required Feynman diagrams we use
Qgraf~\cite{Nogueira:1991ex}. The Qgraf output is easily manipulated
using Form~\cite{Vermaseren:2000nd} to write the amplitude in the form
\beq
A(1,2,3,4,5,6)=\sum_{N=2}^6\sum_{M=0}^N 
K_{\mu_1\cdots\mu_M}(p_1,\epsilon_1;\ldots;p_6,\epsilon_6) 
I_N^{\mu_1\cdots\mu_M}(p_1,\ldots,p_6) \, ,
\eeq
where the kinematic tensor $K$ depends on the purely four-dimensional
external vectors and contains all the particle and process
information. The $N$-point tensor integrals of rank $M$ are defined in
$D$ dimensions as
\beq
I_N^{\mu_1\cdots\mu_M}(p_1,\ldots,p_6)=
\int \frac{d^Dl}{i \pi^{D/2}} \frac{l^{\mu_1}\ldots l^{\mu_M}}{d_1d_2 \ldots d_N}, 
\;\;\; d_i \equiv (l+q_i)^2,\;\;\; q_i \equiv \sum_{j=1}^i p_j\,,
\eeq
and can be evaluated semi-numerically.

For $N\leq 4$ we use the method of
\cite{Giele:2004ub,Giele:2004iy,Ellis:2005zh} which we already
developed, tested and used in the calculation of $\mbox{H} + 4\ 
\mbox{partons}$ at one-loop~\cite{Ellis:2005qe}.  In general, the
basis integrals will contain divergences in $\epsilon=(4-D)/2$ from
soft, collinear and ultraviolet divergences and the answer returned by
the semi-numerical procedure will be a Laurent series in inverse
powers of $\epsilon$.

For the five~(six)-point tensor integrals the method we use relies on
the completeness (over-completeness) of the basis of external momenta
for a generic phase space point.  We therefore use a technique for
tensor reduction which generalizes the methods of
ref.~\cite{vanNeerven:1983vr,vanOldenborgh:1989wn}.  This technique is
valid as long as the basis of external momenta is
complete\footnote{For exceptional momentum configurations (such as
  threshold regions or planar event configurations) this is not the
  case.  Exceptional configurations can be treated using a
  generalization of the expanded relations proposed in
  refs.~\cite{Giele:2004ub,Ellis:2005zh}. This is beyond the scope of
  this paper.}. Assuming we have a complete basis of external momenta
we can select a set of 4 momenta $\{p_{k_1},p_{k_2},p_{k_3},p_{k_4}\}$
which form the basis of the four-dimensional space.  We can then
decompose the loop momentum
\begin{equation}
l^\mu=\sum_{i=1}^4 l\cdot p_{k_i} v_{k_i}^\mu
=V^\mu+\frac{1}{2}\sum_{i=1}^4 \left(d_{k_i}-d_{k_i-1}\right) v_{k_i}^\mu\,,
\end{equation}
where the $v_{k_i}$ are defined as linear combinations of the basis
vectors
\begin{equation}\label{axial}
v^{\mu}_{k_i} = \sum_{j=1}^4 [G^{-1}]_{ij} p^\mu_{k_j}, \;\;\; G_{ij} =p_{k_i} \cdot p_{k_j}\,,
\end{equation}
where  $G$ is the Gram matrix and 
\begin{equation}
V^\mu=-\frac{1}{2}\sum_{i=1}^4 (r_{k_i}-r_{k_i-1})v^\mu_{k_i},\;\;\;
r_k=q_k^2\,. 
\end{equation}
With this relation it is now easy to reduce an $N$-point function of
rank $M$ to a lower rank $N$-point function and a set of lower rank
$(N-1)$-point functions
\beq
I_N^{\mu_1\cdots\mu_M}=I_N^{\mu_1\cdots\mu_{M-1}}V^{\mu_M}
+\frac{1}{2}\sum_{i=1}^4\left(I_{N,k_i}^{\mu_1\cdots\mu_{M-1}}-I_{N,k_i-1}^{\mu_1\cdots\mu_{M-1}}\right)v_{k_i}^{\mu_M}\,,
\eeq
where $I_{N,j}$ is a $(N-1)$-point integral originating from $I_N$
with propagator $d_j$ removed.  More explicitly, choosing without loss
of generality the base set $\{p_1,p_2,p_3,p_4\}$, we get
\beqa 
\lefteqn{I_N^{\mu_1\cdots\mu_M}(p_1,p_2,p_3,p_4,p_5,\ldots,p_N)=
I_N^{\mu_1\cdots\mu_{M-1}}(p_1,p_2,p_3,p_4,p_5,\ldots,p_N) V^{\mu_M}(p_1,p_2,p_3,p_4)}
\nonumber\\&+&\frac{1}{2}
\left(I_{N-1}^{\mu_1\cdots\mu_{M-1}}(p_1+p_2,p_3,p_4,p_5,\ldots,p_N)
-I_{N-1}^{\mu_1\cdots\mu_{M-1}}(p_2,p_3,p_4,p_5,\ldots,p_N)\right)
\nonumber\\&&\times
v_1^{\mu_M}(p_1,p_2,p_3,p_4)
\nonumber\\&+&\frac{1}{2}
\left(I_{N-1}^{\mu_1\cdots\mu_{M-1}}(p_1,p_2+p_3,p_4,p_5,\ldots,p_N)
-I_{N-1}^{\mu_1\cdots\mu_{M-1}}(p_1+p_2,p_3,p_4,p_5,\ldots,p_N)\right)
\nonumber\\&&\times
v_2^{\mu_M}(p_1,p_2,p_3,p_4)
\nonumber\\&+&\frac{1}{2}
\left(I_{N-1}^{\mu_1\cdots\mu_{M-1}}(p_1,p_2,p_3+p_4,p_5,\ldots,p_N)
-I_{N-1}^{\mu_1\cdots\mu_{M-1}}(p_1,p_2+p_3,p_4,p_5,\ldots,p_N)\right)
\nonumber\\&&\times
v_3^{\mu_M}(p_1,p_2,p_3,p_4)
\nonumber\\&+&\frac{1}{2}
\left(I_{N-1}^{\mu_1\cdots\mu_{M-1}}(p_1,p_2,p_3,p_4+p_5,\ldots,p_N)
-I_{N-1}^{\mu_1\cdots\mu_{M-1}}(p_1,p_2,p_3+p_4,p_5,\ldots,p_N)\right)
\nonumber\\&&\times
v_4^{\mu_M}(p_1,p_2,p_3,p_4)\,. 
\nonumber\\
\eeqa

For example, applying this relation repeatedly to the tensor six-point
integrals we will be left with the scalar six-point integral and
five-point tensor integrals. The five-point tensor integrals can be
reduced using the same technique. Subsequently we can use the method
of~\cite{Giele:2004ub,Giele:2004iy,Ellis:2005zh} to further
numerically reduce all remaining integrals to the basis of scalar 2-,
3- and 4-point integrals. This procedure turns out to be efficient and
straightforward to implement numerically.

\section{Comparison with the literature}
Since we have directly calculated the loop amplitudes with internal
gluons and fermions we can easily obtain the result for QCD with an
arbitrary number $n_f$ of flavours of quarks,
\begin{equation}
 {A}^{\rm QCD} = A^{[1]} + \frac{n_f}{N} A^{[1/2]}\, .
\end{equation}
However since the analytic calculations in the literature are
presented in terms of supersymmetric theories we need to re-organize
our results to compare with other authors.
\subsection{Supersymmetry}
Since we have calculated the amplitudes with massless spin $1$, spin
$1/2$ and spin $0$ particles in the internal loop we can combine our
results as follows
\begin{eqnarray}
{A}^{{\N}=4}&=&A^{[1]}+4 A^{[1/2]}+3 A^{[0]}\, , \\
{A}^{{\N}=1}&=& A^{[1/2]}+A^{[0]}.
\end{eqnarray}
${A}^{{\N}=4}$, so constructed, describes an amplitude where the full
supersymmetric ${\N}=4$ multiplet runs in the loop, and ${A}^{{\N}=1}$
denotes the contribution from an ${\N}=1$ super-multiplet running in
the loop.

In analytic calculations the intention is to proceed in the opposite
direction.  Amplitudes with multiplets of supersymmetric Yang-Mills in
internal loops have much improved ultra-violet behavior and are
four-dimensional cut-constructible. For this reason, all of these
supersymmetric amplitudes have been calculated and most have been
presented in a form suitable for numerical evaluation.  As far as
six-gluon amplitudes with scalars in the loop, ${A}^{[0]}$, are
concerned three of the needed eight independent helicity amplitudes
have been published so far. Only in the helicity combinations where
all contributions are known can one reconstruct the ingredients needed
for QCD amplitudes
\begin{eqnarray}
{A}^{[1]}&=&{A}^{{{\N}}=4}-4{A}^{{\N}=1}+{A}^{[0]} \, ,\\ 
{A}^{[1/2]}&=& {A}^{{\N}=1}-{A}^{[0]}\, .
\end{eqnarray}

\subsection{Numerical results}
As a preparatory exercise we performed a check of the four- and
five-point gluon one-loop amplitudes.  We found agreement with the
literature~\cite{Ellis:1985er,Kunszt:1993sd,Bern:1993mq}.

We now turn to the amplitude for six-gluons which is the main result
of this paper.  Our numerical program allows the evaluation of the
one-loop amplitude at an arbitrary phase space point and for arbitrary
helicities. For a general phase space point it is useful to re-scale
all momenta so that the momenta of the gluons, (and the elements of
the Gram matrix), are of $O(1)$ before performing the tensor
reduction. Without loss of generality we can assume that this has been
done.

To present our numerical results we choose a particular phase space
point with the six momenta $p_i$ chosen as follows, $(E,p_x,p_y,p_z)$,
\begin{eqnarray}
\label{specificpoint}
     p_1 & = & \frac{\mu}{2} (-1,  +\sin\theta, +\cos\theta \sin\phi, +\cos\theta \cos\phi ), \nonumber \\
     p_2 & = & \frac{\mu}{2} (-1,  -\sin\theta, -\cos\theta \sin\phi, -\cos\theta \cos\phi ), \nonumber \\
     p_3 & = & \frac{\mu}{3} (1,1,0,0), \nonumber \\
     p_4 & = & \frac{\mu}{7} (1,\cos\beta,\sin\beta,0), \nonumber \\
     p_5 & = & \frac{\mu}{6} (1,\cos\alpha \cos\beta, \cos\alpha \sin\beta,\sin\alpha), \nonumber \\
     p_6 & = & -p_1-p_2-p_3-p_4-p_5\, ,
\end{eqnarray}
where $\theta= \pi/4,\phi= \pi/6,\alpha= \pi/3,\cos \beta= -7/19$.
Note that the energies of $p_1$ and $p_2$ are negative and $p_i^2=0$.
In order to have energies of $O(1)$ we make the choice for the scale
$\mu=n=6$~[GeV].  As usual $\mu$ also denotes the scale which is used
to carry the dimensionality of the $D$-dimensional integrals.  The
results presented contain no ultraviolet renormalization.

Analytic results require the specification of eight helicity
combinations: all other amplitudes can be obtained by the parity
operation or cyclic permutations.  We choose these eight combinations
to be the two finite amplitudes ($++++++,-+++++$), the maximal
helicity violating amplitudes ($--++++,-+-+++,-++-++$), and the
next-to-maximal helicity violating amplitudes
($---+++,--+-++,-+-+-+$).  These eight amplitudes would not be
sufficient for a numerical evaluation, but the numerical approach
allows the evaluation of any helicity configuration at will.

In Table~\ref{tableneq4} we give results for a particular colour
sub-amplitude ${A}^{{\N}=4}(1,2,3,4,5,6)$ for the above eight choices
of the helicity. An overall factor of $i \cg$ has been removed from
all the results in the Tables~\ref{tableneq4}, \ref{tableneq1}, and
\ref{tablescalar}
\begin{equation}
\cg = {(4 \pi)^\eps \over 16 \pi^2 }
{\Gamma(1+\eps)\Gamma^2(1-\eps)\over\Gamma(1-2\eps)}\ .
\label{cgdef}
\end{equation}
The results for the ${\N}=4$ amplitudes depend on the number of
helicities of gluons circulating in internal loops.  
For a recent description 
of regularization schemes see, for example, ref.~\cite{Bern:2002zk}. 
Our results are
presented in the 't Hooft-Veltman scheme. 
The translation to the four-dimensional helicity scheme is immediate
\begin{equation}
{A}^{{\N}=4}_{\rm FDH} = {A}^{{\N}=4}_{\rm t-HV} + \frac{\cg}{3}
{A}_{\rm tree}\,. 
\label{THVtoFDH}
\end{equation}
Note that analytic results from the literature are quoted in the
four-dimensional helicity scheme, which respects supersymmetry.  These
results have been translated to the 't Hooft-Veltman scheme using
Eq.~(\ref{THVtoFDH}) before insertion in our tables.

\begin{scriptsize}
\TABLE{
\begin{tabular}{|c|c|c|c|c|}
\hline
Helicity & $1/\epsilon^2$ & $1/\epsilon$ & 1 &[Ref]/(Eq.\#) \\
\hline
$++++++$ & $0$                         &  $0$                          & $0$  &   \\
$++++++$ & $(-1.034+i~2.790 ) 10^{-8}$&$  (-9.615+i~3.708 ) 10^{-8}$&$  -(0.826+i~2.514) 10^{-7}$ &  [SN-A]   \\
\hline
$-+++++$ & $0$                        &  $0$                        & $0$   &   \\ 
$-+++++$ & $(1.568+ i~2.438) 10^{-8}$ & $ (-0.511 +i~1.129) 10^{-7}$&$ -(3.073+i~0.1223) 10^{-7}$  &  [SN-A]   \\ 
\hline
\hline
$--++++$ & $-161.917+i~54.826 $         & $  -489.024-i~212.415 $      & $ -435.281-i~1162.971 $  & \cite{Bern:1994zx}/(4.19)\\
$--++++$ & $(-0.933 +i~1.513) 10^{-8} $ & $ -(7.655+i~0.440)10^{-8}  $ & $ -(-0.221+i~1.834)10^{-7} $  & [SN-A]  \\
\hline
$-+-+++$ & $ -33.024 + i~44.423 $      & $  -169.358 + i~33.499 $     & $ -330.119 -i~229.549 $ & \cite{Bern:1994zx}/(4.19) \\
$-+-+++$ & $(-7.542+i~0.939) 10^{-8} $ & $ -(1.157 +i~0.363)10^{-8} $ & $  -(3.474 +i~2.856)10^{-8} $ &  [SN-A] \\
\hline
$-++-++$ & $ -0.5720 - i~3.939 $       & $ 6.929 - i~10.302  $         & $ 28.469 -i~5.058 $      & \cite{Bern:1994zx}/(4.19) \\
$-++-++$ & $(-2.279 +i~1.803)10^{-8} $ & $  -(1.176 +i~0.399)10^{-7} $ & $ (0.054-i~3.307)10^{-7} $      &  [SN-A] \\
\hline
\hline
$---+++$ & $ -6.478 -i~10.407 $       & $ 6.825 -i~37.620  $       & $ 75.857 - i~47.081 $            & \cite{Bern:1994cg}/(6.19) \\
$---+++$ & $ (2.686-i~1.668)10^{-8} $ & $ (1.232+i~0.554)10^{-7} $ & $ (0.020+i~3.334 )10^{-7} $      &  [SN-A]  \\
\hline
$--+-++$ & $ 14.074-i~22.908 $         & $ 80.503- i~23.464 $         & $  169.047 + i~93.601 $       & \cite{Bern:1994cg}/(6.24) \\
$--+-++$ & $ -(1.619+i~0.943)10^{-8} $ & $  -(1.030+i~8.234)10^{-8} $ & $ (1.560 -i~0.801)10^{-8} $    &  [SN-A]  \\
\hline
$-+-+-+$ & $ 13.454+i~13.177 $        & $ 3.495+i~58.632 $           & $ -88.32+i~103.340 $             & \cite{Bern:1994cg}/(6.26)  \\
$-+-+-+$ & $ (1.045-i~0.113)10^{-9} $ & $ (-0.772+i~1.652))10^{-8} $ & $ (-7.795+i~7.881))10^{-8} $      & [SN-A]  \\
\hline
\hline
\end{tabular}
\caption{$\N$=4 color ordered sub-amplitudes evaluated at the specific point, Eq.~(\ref{specificpoint}).
  The results are given in the 'tHooft-Veltman regularization scheme.
  [SN-A] means the difference between the semi-numerical result and
  the analytical one.}
\label{tableneq4}
}

\end{scriptsize}

\begin{scriptsize}
\TABLE{
\begin{tabular}{|c|c|c|c|c|}
\hline
Helicity &  $1/\epsilon^2$ & $1/\epsilon$ & 1 &[Ref]/(Eq.\#) \\
\hline
$++++++$ &  0 & 0  &  0 & \\
$++++++$ & $(-3.470+i~9.320) 10^{-9}$&$(-3.226+i~1.253) 10^{-8}$&$ -(3.899+i~8.969) 10^{-8}$ & [SN-A] \\
\hline
$-+++++$ &   0 & 0 &  0 & \\
$-+++++$ & $(5.228+i~8.127) 10^{-9}$&$(-1.678+i~3.775) 10^{-8} $&$ -(1.013+i~0.2066) 10^{-7} $ & [SN-A] \\
\hline
\hline
$--++++$ &  0                        & $26.986-i~9.1376$              & $101.825-i~52.222$ &  \cite{Bern:1994cg}/(5.9)\\
$--++++$ &$(-3.297+i~5.194) 10^{-9}$ & $ -(-2.104+i~0.344) 10^{-8}  $ & $(0.949 -i~4.895) 10^{-8} $ & [SN-A] \\ 
\hline                                       
$-+-+++$ &  $0$                                 & $ 5.504-i~7.404 $ & $ 21.811-i~29.051 $ &  \cite{Bern:1994cg}/(5.12)\\
$-+-+++$ & $(-1.847 + i~0.8566) 10^{-10} $ & $ -(6.141+i~4.633 ) 10^{-10}  $ & $ (3.095+i~2.138) 10^{-7}   $ &  [SN-A] \\
\hline                                        
$-++-++$ &  $0$                        & $0.09533+i~0.6565$           & $ -2.183+i~3.260 $ & \cite{Bern:1994cg}/(5.12)\\
$-++-++$ &  $(-7.599+i~6.018) 10^{-9}$ & $ -(3.929+i~1.304)10^{-8}  $ & $(0.008-i~1.100)10^{-7}  $ &  [SN-A] \\
\hline
\hline
$---+++$ &  $0$                       & $1.080 +i~1.735$           & $ 0.722+i~5.285$   &  \cite{Bidder:2004tx}/(9) \\
$---+++$ &  $(8.965-i~5.555) 10^{-9}$ & $(4.107 +i~1.858)10^{-8} $ & $ (0.002+i~1.114)10^{-7} $ &  [SN-A] \\
\hline                                           
$--+-++$ &  $0$                        & $-2.346+i~3.819$ &                    &  \cite{Britto:2005ha}/(5.4,2.3)\\
$--+-++$ &  $(-5.351-i~2.825) 10^{-9}$ & $-2.346+i~3.819$ & $-2.238+i~17.687$  &  [SN]  \\
\hline                                           
$-+-+-+$ &  $0$                             & $-2.242-i~2.196$ &              &  \cite{Britto:2005ha}/(5.13,2.3)\\
$-+-+-+$ &  $(1.124-i~0.2060) 10^{-10}$ & $-2.242-i~2.196$ & $-1.721-i~7.433$ &  [SN]  \\
\hline
\hline
\end{tabular}
\caption{$\N$=1 color ordered sub-amplitudes evaluated at the specific point, Eq.~(\ref{specificpoint}).
  [SN] means that the result is obtained using our semi-numerical
  code, while [SN-A] denotes the difference between the semi-numerical
  result and the analytical one.}
\label{tableneq1}
}

\end{scriptsize}

In Table~\ref{tableneq1} we give results for the colour sub-amplitudes
${A}^{{\N}=1}(1,2,3,4,5,6)$ for the same eight helicity choices and
where possible compare with analytical results.~\footnote{In
  Eq.~(5.16) of ref.~\cite{Bern:1994cg} for the degenerate case
  m=j-1=2 one has $\hat{{\cal C}}_m = \{j+1, \ldots, n-1 \} $, as can
  be seen from Fig.~8 of this same paper.  This point has also been
  made in ref.~\cite{Cachazo:2004zb}. }  
Note that because of the relation
\begin{equation}
{A}^{{\N}=1}|_{\rm singular} = \frac{\cg}{\epsilon}  \Atree\, ,
\end{equation}
the column giving the single pole can as well be considered as a
listing of the results for the colour-ordered sub-amplitudes at tree
graph level (stripped only of the overall factor of $i$).

We note that for two of the helicity amplitudes $--+-++$ and $-+-+-+$
we were unable to evaluate the analytic results numerically. This was
due to the fact that calculating the residue of certain poles as
required by the formula in ref.~\cite{Britto:2005ha}, resulted in zero
value denominators of sub-expressions\footnote {We thank the authors
  of ref.~\cite{Britto:2005ha} for confirming that there are problems
  with the numerical evaluation of the formula for these amplitudes in
  their paper.}.

\begin{scriptsize}
\TABLE{
\begin{tabular}{|c|c|c|c|c|}
\hline
Helicity &  $1/\epsilon^2$ &  $1/\epsilon$ & 1 & [Ref]/(Eq.\#) \\
\hline
$++++++$ &  $0$                          & $0$                          & $  (4.867 + i~2.092)  10^{-1}$&\cite{Bern:2005ji}/(4.3)\\ 
$++++++$ &  $(3.672 +i~9.749)  10^{-9} $ & $(-3.404 + i~1.238)  10^{-8}$& $ -(3.016+ i~9.169) 10^{-8} $& [SN-A] \\
\hline
$-+++++$ &   0                           & 0                             & $-3.194 + i~0.6503  $  &  \cite{Bern:2005ji}/(4.10)\\ 
$-+++++$ &   $(5.921 +i~8.411)  10^{-9}$ & $(-1.606 +i~4.051)  10^{-8} $ & $ -(1.086 +i~0.038) 10^{-7}   $  & [SN-A] \\
\hline
\hline
$--++++$ &  $0$                                & $8.995-i~3.046 $&  {$43.089-i~20.288 $} &\cite{Bern:2005cq}/(4.27,4.28) \\ 
$--++++$ & $(1.280 + i~0.002) 10^{-8}$ & $(2.768+i~4.232) 10^{-8} $     & $ (-1.004+i~0.955)10^{-7}   $ & [SN-A]  \\ 
\hline
$-+-+++$ & $(1.045-i~0.580) 10^{-8}$   & $1.835-i~2.468 $               & $9.752-i~11.791$ & [SN] \\ 
\hline
$-++-++$ & $(-7.791+i~6.717) 10^{-9}$  & $3.178\cdot 10^{-2}+i~0.2188 $ & $-1.447+i~0.1955$  & [SN]  \\ 
\hline
\hline
$---+++$ & $(8.934-i~5.359) 10^{-9}$   & $0.3599+ i~0.5782$             & $ 0.5617+i~5.8166$ & [SN] \\ 
\hline
$--+-++$ & $(0.1016 +i~1.276) 10^{-8}$ & $  -0.7819 +i~1.273 $          & $  -0.6249+i~6.552$  & [SN]  \\ 
\hline
$-+-+-+$ & $(1.065- i~0.5417) 10^{-8}$ & $ -0.7475-i~0.7321 $           & $  -1.298 - i~3.255$ & [SN]  \\ 
\hline
\hline
\end{tabular}
\caption{One loop six gluon colour ordered sub-amplitudes with a scalar loop 
  evaluated the specific point Eq.~(\ref{specificpoint}).  [SN] means
  that the result is obtained using our semi-numerical code, while
  [SN-A] denotes the difference between the semi-numerical result and
  the analytical one.}
\label{tablescalar}
}

\end{scriptsize}

Lastly in Table~\ref{tablescalar} we give results for the colour
sub-amplitudes $A^{[0]}(1,2,3,4,5,6)$ for scalar gluons, for the same
eight helicity choices.\footnote{In ref.~\cite{Bern:2005cq} [v1-v3]
  the definition of $F_f$ has an overall sign missing, a typographical
  error not present in the original calculation of the $\N$ = 1 term
  in ref.~\cite{Bern:1994cg}.}
For all amplitudes for which no analytic result exists, we checked the
gauge invariance of the amplitudes by changing the gluon polarization.
The gauge invariance was obeyed with a numerical accuracy of ${\cal
  O}\left(10^{-8}\right)$. To evaluate a single colour-ordered
sub-amplitude for a complex scalar took 9 seconds on a 2.8GHz Pentium
processor. To evaluate the complete set of 64 possible helicities will
be less than 64 times longer, because the scalar integrals are stored
during the calculation of the first amplitude are applicable to all
other configurations with the same external momenta.

\section{Conclusions}
In this paper we have presented numerical results which demonstrate
that the complete one-loop amplitude for six-gluon scattering is now
known numerically.
By forming multiplets of SUSY Yang Mills in the internal loops, 
we were able compare with most of the known analytic results.
In addition, we have presented numerical results for amplitudes which
are currently completely unknown. Note that the analytic and
semi-numerical results are complementary.  The hardest piece to
calculate analytically is the scalar contribution $A^{[0]}$, which is
the easiest for the semi-numerical approach. Thus it is possible that
a numerical code involving both semi-numerical and analytic results
will be the most efficient and expedient.  Our results demonstrate the
power of the semi-numerical method, which can supplant the analytic
method where it is too arduous and provide a completely independent
check where analytic results already exist.

After inclusion of the one-loop corrections to the other parton
subprocesses involving quarks it would be possible to proceed to a NLO
evaluation of the rate for four jet production.  We intend to use
these methods to calculate NLO corrections to other processes which we
consider to be of more pressing phenomenological interest.

\section*{Acknowledgements}
We would like to thank Zvi Bern, Lance Dixon and David Kosower for
providing helpful comments on the draft of this manuscript. We also
acknowledge useful discussions with John Campbell, Vittorio Del Duca
and Fabio Maltoni.

\end{document}